\documentclass{emulateapj}
\usepackage{apjfonts,natbib,epsfig,color}
\usepackage{amsmath}

\usepackage{graphicx}
\newcommand{\beq}{\begin{equation}}
\newcommand{\beqa}{\begin{eqnarray}}
 \newcommand{\eeq}{\end{equation}}
\newcommand{\eeqa}{\end{eqnarray}}

\newcommand{\lsim}{\lesssim}

\newcommand{\lmk}{\left(}
\newcommand{\rmk}{\right)}

\shorttitle{}
\shortauthors{}

\begin{document}

\title{The Search for Extra-Galactic Intelligence Signal Synchronized with a Binary Neutron Star Merger}

\author{Yuki Nishino and Naoki Seto}

\affil{Department of Physics, Kyoto University, Kyoto 606-8502, Japan}

\begin{abstract}

We discuss  the possibility of receiving a radio signal from extra-Galactic intelligence, around the time when  we observe a binary neutron star  merger in their galaxy. High-precision measurements of the binary parameters allow them to send the signal  $\sim10^4$ years before they themselves observe the merger signal.  Using the SKA,  we might receive $\sim10^4$ bits of data, transmitted from 40 Mpc distance with the output power of $\sim1$ TW. 
We also discuss related topics for GW170817 and  mention potential roles of future gravitational wave detectors in relation to this transmission scheme.  

\end{abstract}


\section{INTRODUCTION}
Obtaining evidence of extraterrestrial intelligence (ETI taken as plural) is a fascinating  long-term objective in astronomy. At present, considering the expected signal strength, signatures of Galactic ETI  are the primary observational target. 
In fact, since the pioneering work by Drake (1961), many "search for ETI" (SETI) projects have weighted Galactic systems (Tarter et al. 1980; Horowitz \& Sagan 1993; Steffes \& DeBoer 1994; Rampadarath et al. 2012; Siemion et al. 2013). 
Though the Galactic searches have not succeeded yet (see also Kraus 1979), continuous efforts could be highly worthwhile, and currently operational and forthcoming observational facilities such as GBT, Parkes, Arecibo, VLA, ATCA, GMRT, MWA, MeerKAT, ATA, LOFAR, FAST, and SKA are useful for this endeavor.

In the meantime, there is still the possibility that  Galactic ETI signals (both active and passive types) might be difficult to detect. This might be caused by the following reasons. (i) The total number of Galactic civilizations might be small, e.g. because of their exceedingly short lifetimes (Ulmschneider 2006).  (ii) From the security perspective, considering the travel time within the Galaxy, ETI might attempt to cloak their techno-signatures or even bio-signatures, rather than  purposefully transmitting signals to attract other Galactic civilizations (Kipping \& Teachey 2016).

Therefore, extra-Galactic SETI could be a  fruitful option, complementary to the Galactic searches (see also  Zackrisson et al. 2015). 
Note that, for a purposeful signal transmission outward from their galaxy,  because of the required travel distances, the security issue would be considerably mitigated. However, at the same time, due to the dilution, detecting of an extra-Galactic signal would be more difficult than detecting Galactic signals. 
This difficulty would also be understood also by the extra-Galactic ETI, and they would carefully arrange the timing and direction of their intended signal transmissions, considering the mutually obvious points in the strategy space (Wright 2018), namely the Schelling point in game theory (Schelling 1960).

In this relation, various astrophysical systems have been proposed as a potential tool for synchronized signal transmission (e.g., periodic orbital motions, Pace \& Walker 1975; supernovae, Tang 1976; stellar flares, Siebrand 1982; gamma-ray bursts, Corbet 1999). Here,  the timing accuracy and the relative separation between the three points (the sender, the receiver, and the adopted astronomical system)  are the key elements, limiting the range of the actual application.

On 2017 August 17, the LIGO-Virgo network detected an inspiral gravitational wave (GW) from a binary neutron star (BNS) system (GW170817; Abbott et al. 2017a). Soon after this highly energetic event, its host galaxy NCG4993 was identified by electromagnetic observations, at a distance of $\sim 40$ Mpc (Abbott et al. 2017b). It is an elliptical galaxy with a stellar mass similar to the Milky Way but a somewhat ($\times 1/3$) smaller spatial size (Im et al. 2017).
Including GW170817, the volume-averaged BNS merger rate was  estimated, and its central value is  $R \sim 1.5\times 10^{-6}  {\rm Mpc^{-3} yr^{-1}}$ (Abbott et al. 2017a). 
This rate is consistent with that estimated from the known  BNSs in our Galaxy (Abadie et al. 2010). These Galactic BNSs contain recycled pulsars as one of the binary components, allowing high-precision measurements of the binary parameters.  If ETI can detect the recycled pulsar of an inspiraling BNS system in their galaxy, they can also make a high-precision measurement for the quantities required for the signal transmission  to be synchronized with the BNS merger.  
This synchronization scheme seems reasonable also to us as a signal receiver, and thus could be  on the Schelling point mentioned earlier.

In this paper, we discuss the prospects of this scheme. Given the impacts of the ETI signal detection, we primarily regard ourselves as a receiver. For technological feasibility, a receiver is assumed to have a sensitivity up to the phase 2 of the SKA (around 2030), and we  specifically  discuss our present situation in relation to GW170817.  For a sender, we mainly presume a technological level that might  be realized in 50-100 years on the Earth, taking into account the future plans currently under discussion. 
We also mention the potential roles of future GW detectors such as the Einstein telescope, LISA, BBO, and DECIGO.

To simplify our arguments, we deal with a model in which  our universe is composed by  single-species galaxies  equivalent to the Milky Way. This is based on the fact that $\sim 50\%$ of the B-band luminosity density is contained in galaxies more luminous than the Milky Way. Below, we use the corresponding number density of the galaxies $n_{\rm M}\sim 10^{-2} \rm  Mpc^{-3}$ (Phinney 1991; Kalogera et al. 2001) and the BNS merger rate $R_{\rm M}\equiv R/n_{\rm M}=1.5 \times 10^{-4}  {\rm yr^{-1}}$ in each galaxy.  For high-precision measurements, the pulsar beam should intersect with the ETI's line of sight. Given the beaming fractions of pulsars, the number of BNS systems available for the synchronization is reduced by a correction factor $\eta=6$ (following Kalogera et al. 2001).  We adopt the available BNS merger rate $ \eta^{-1}R_{ \rm M} \sim 3 \times 10^{-5} { \rm yr}^{-1} $ hereafter. We  assume that the mass of a neutron star is $\sim 1.4 M_\odot$.

\section{Preset Transmission for Synchronous  Reception}

We consider an intentional signal transmission by extra-Galactic ETI. They are  assumed to utilize a BNS system in their galaxy and adjust the direction and timing of their signal transmissions, so that, at the target galaxy, the arrival time of their signal synchronizes with that of the merger GW signal (see Fig. 1).

We introduce their time coordinate $t_{\rm e}$, and put $t_{\rm e}= t_{\rm e,gw}$ for the moment when the merger GW signal reaches them. We additionally  define $\Delta t_{\rm e}\equiv t_{\rm e}-t_{\rm e,gw}$ for the relative time. 
 For the synchronization, they need to select the azimuthal angle $\theta$  of the shooting direction (see Fig.1) as a function of the time $\Delta t_{\rm e}$
\beq
\Delta t_{\rm e}=-\frac{l}{c} (1+\cos\theta). \label{th}
\eeq
Here, $l$ is the distance  between the BNS and the ETI.
From eq.(\ref{th}), the signal can be transmitted  during the time interval $-2l/c\le \Delta t_{\rm e}\le 0$.  Correspondingly, the target sky direction moves from $\theta=0$ to $\pi$, increasing the completed sky area at the constant rate $2\pi c/l$.

Note that the characteristic time $l/c=3\times 10^4(l/10{\rm kpc})\ $yr  is similar to the expected interval $\eta \,R_{\rm M}^{-1}$ of the available BNS merger in their galaxy.
Therefore, at any time, the ETI have 
  O(1) BNSs utilizable for the synchronization, unless they could  find only a tiny fraction of short-period BNSs.  We should also mention that, at $3\times 10^4$ yrs before the merger,  a BNS has a GW frequency of $f \sim 4$ mHz and would be easily identified with their LISA-like detector, promoting follow-up radio observations. \footnote{The radio observations would be more suitable for the present purpose. However, in some cases, black hole binaries might be also used for the synchronized signal transmissions (possibly in the gamma-ray band).}

Here, we briefly discuss the pulsar timing analysis for such a short-period BNS.  To this end, we temporarily use PSR J0737-3039 (orbital period $P=2.4$ hr) as a representative example, since this BNS has the largest contribution to the estimated Galactic BNS merger rate (Abadie et al. 2010).   On a nearly edge-on orbit (inclination angle $\sim 89^\circ$),  this binary has two pulsars: the recycled one A, (the current spin period $P_{\rm A}=22.7$ ms),  and the younger one B, ($P_{\rm B}=2.77$ s) (Burgay et al. 2003; Lyne et al. 2004).  In $\sim$ 85 Myr, its orbital separation will decrease down to $\sim1.3\times10^5$ km (corresponding to the orbital period $\sim$ 500 s and the GW frequency $\sim4$mHz) with the residual eccentricity $\sim 4\times 10^{-3}$. From the current spin-down rates (ignoring the pulsar deathline), the two spin periods of the evolved BNS are estimated to be  $P_{\rm A}=27$ ms and $P_{\rm B}=5.1$ s. Thus, the orbital separation  is now smaller than the light cylinder of B, $cP_{\rm B}/(2\pi)\sim 2\times 10^5$ km.  However,  because of the $\sim 50,000$ times stronger wind from A, the magnetosphere of B would be highly deformed with the bow shock distance $\sim 3\times10^4$ km (based on Arons et al. 2005).  Therefore, similar to the currently presumed situation (Lyne et al. 2004), the pulsar activity of A in itself would not be largely affected by that of B.
Meanwhile, we should note the possibility that the observed pulse profile of A might be modified by B, depending on the orbital phase. For a typical orbital inclination angle (e.g. $30^\circ$), the modification might be smaller than what currently happens for the nearly edge-on system, PSR J0737-3039 (see, e.g., Kramer et al. 2006). In any case, to suppress the systematic errors, careful analysis should be performed by comparing independent parameter estimation methods (the apsidal precession, the Shapiro time delay, GW observation etc).

 In reality, at the target galaxy, the synchronization would not be adjusted perfectly, because of the estimation error of related parameters. There would be a certain difference between the  arrival time of the merger GW signal and that of the synchronized artificial  signal.  The ETI would carefully control this time difference.  If the ETI attempt to send  a single short signal  with a  mismatch time of less than $\delta t$, they  need to estimate  $\Delta t_{\rm e}$  and $l/c$ at the same accuracy. 
From our current standpoint as a fledgling GW observer, a mismatch time $\delta t \lsim  1$ years would be  desirable, given the extensive  year-long observations for the  GW170817 afterglow.

Next, we discuss the accuracy of the parameter estimation. The gravitational orbital evolution depends strongly on the GW frequency $f$ as  ${\dot f}\propto f^{11/3}$.
The remaining time $\Delta t_{\rm e}$ before the BNS merger is given by 
\beq
\Delta t_{\rm e}\simeq -\frac38 \frac{f}{\dot f} \simeq -\frac{15c^5}{768 (2\pi)^{8/3} G^{5/3}} f^{-8/3} m_{\rm c}^{-5/3}  \label{time}
\eeq
 where we dropped the well-known but somewhat complicated  dependence on the orbital eccentricity (Peters 1964).
Here, $m_{\rm c}$ is the chirp mass of  the BNS and is given by the individual masses as $m_{\rm c}\equiv (m_1 m_2)^{3/5}(m_1+m_2)^{-1/5}$. Therefore, to realize the mismatch  $\delta t<1$ years for the typical remaining time $\Delta t_{\rm e}\sim -3\times 10^4$ years, both the chirp  mass $m_{\rm c}$ and the frequency $f$ should be estimated at 5 digits (smaller digits for $e\ll 1$).

Meanwhile, with respect to nearly monochromatic binaries,  a LISA-like GW detector has a simple scaling relation for the parameter estimation error $\Delta {\dot f}=\zeta \ t_{\rm obs}^{-2} \ (S/N)^{-1}\propto t_{\rm obs}^{-5/2} $ with the observation period $t_{\rm obs}$, the signal-to-noise ratio $S/N$ and a numerical coefficient $\zeta=O(1)$ (see e.g. Takahashi \& Seto 2002).  For $t_{\rm obs}=30$ years and $S/N=100$, we can reach  the level $\Delta {\dot f}/{\dot f}\sim  10^{-5}$ for a BNS at $f\sim4$mHz. For a radio pulsar timing analysis, we can expect a similar scaling relation $\Delta {\dot f}$ with a much better overall accuracy (for a higher $S/N$).
Therefore, given their advanced technology, the 5 digit accuracy for $m_{\rm c}$  would not be challenging for the ETI (see, e.g., Weisberg et al. (2010) for a long-term observation  of PSR B1913+16 and Damour \& Taylor (1991) for potential astrophysical effects for $\dot f$).  Similarly, they would be  able to estimate the distance $l$ at the level $c\delta t<0.3$ pc, e.g. by using the long-term kinematical parallax (see Smits et al. 2011 for the prospects with the SKA). They would also appropriately measure the related parameters (e.g. their peculiar velocity and acceleration relative to the BNS).

So far, in relation to the afterglow observation of a BNS merger, a mismatch of $\delta t<1$ years  is considered to be desirable.  However, the situation might be different in the future, e.g.  at the time of  the 100th BNS merger detection. 
The precision goal $\delta t$ of the  ETI (targeting such a routine GW observer) is uncertain, but a smaller value would be preferred (e.g. $\delta t \sim 1$week) for a receiver.  Nevertheless, adding a new twist to the signal transmission scheme, the requirement for the precision would  be considerably relaxed.  For example,  the ETI might send two types of signals. The first one is just  for attracting the attention of receivers and the second one is the main message. Considering the timing error,  they might send the first alert signals repeatedly and intermittently, so that one of the alerts would be delivered close to the merger GW signal.  Then, after a certain time (e.g. $\sim$ 5 years), the ETI would start sending the second signal (main message). This is because, given the magnitudes of the impacts,  the receivers are likely to continuously wait for the follow-up signals for a long period of time, once they detect any kind of artificial signal.

\begin{figure}
 \begin{center}
  \includegraphics[width=85mm]{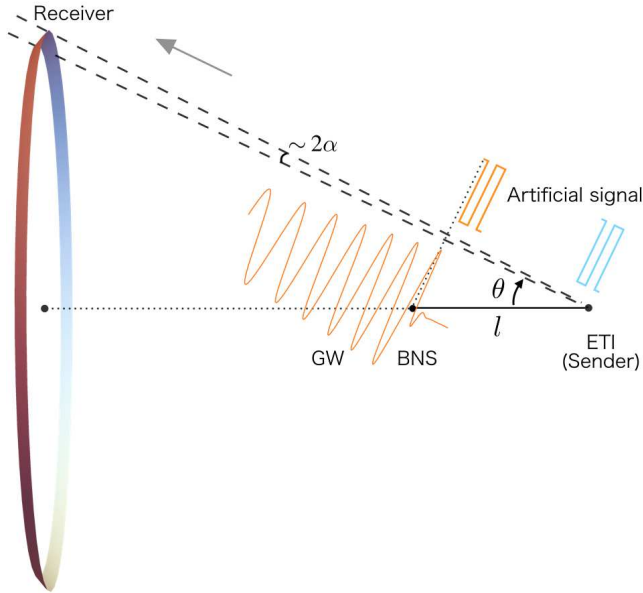}
   \caption{Illustration of the signal transmission scheme, synchronized with an inspiraling BNS system. The ETI transmit an artificial signal (shown with the blue line)  in the direction $\theta$, as a function of time given by eq.(1). Then, after the time $l  \cos\theta / c $,  their signal is synchronized with the merger GW signal  (shown with the orange line). The distance $D$ to the receiver is much longer than the ETI-BNS distance $l$. We use $\alpha$ for the beam width (see \S 4). }
  \label{figure:fig1}
 \end{center}
\end{figure}

\section{The Search for artificial signals}

Considering the successful follow-up observation for GW170817, it is likely that we can identify the host galaxy of a BNS merger at a distance beyond 40 Mpc, and can subsequently search for a synchronized artificial signal coming from somewhere in the galaxy. 
For general GW observers (not only humankind) above 10 Hz, multiple detectors would be used to estimate the binary direction, and the polarization dependence would be averaged. Then the signal-to-noise ratio of the GW detector network depends on the binary inclination $\iota$ as $\propto (1+6\cos^2\iota+\cos^4\iota)^{1/2}$ (Peters 1964).  At the signal transmission, the ETI might take into account this radiation pattern.

Given the practical advantages both for  the senders and receivers, the radio band would be an attractive option for the carrier wave of the signal (Tarter 2001). In particular, the band  around 1-10 GHz is promising,   as originally discussed in Cocconi \& Morrison (1959).  In this band, the synchrotron background form the Galactic plane becomes very small, unlike the lower band $\nu<1$ GHz, but the interstellar medium absorption is weak. Furthermore, on the surface of the Earth, the atmospheric emission and absorption by $\rm H_2O$ and $\rm O_2$ are insignificant,  in contrast to the higher band  $\nu>10$ GHz.   The optimality of the 1-10 GHz band would not be specific to the Earth, but would be widely expected for receivers and senders on habitable planets (Oliver \& Billingham 1971).  
Indeed,  this band has been extensively used for SETI (Tarter 2001). 
However, we should note that, because of the frequency dependence of the scintillation broadening (mentioned shortly) and the beam width (in \S 4), the band above 10 GHz might be selected by the ETI (Benford et al. 2010a; 2010b).

Here, we briefly discuss the detectability of the synchronized narrowband signals in this band. 
To characterize the sensitivity of a radio telescope, there are two fundamental parameters: the collecting area $A$ and the system temperature $T_{\rm sys}$. More specifically, the combination $A/T_{\rm sys}$ is the standard measure for its sensitivity and is often used in the literature.  For example,  in the 1-10 GHz band, the design goal of the  SKA phase 2 is $A/T_{\rm sys}\simeq 10^4 {\rm m^2/K}$ (Dewdney et al. 2009). As for GW170817, radio observations have already been performed by  VLA, GMRT, and ATCA with $A/T_{\rm sys}\sim 100 {\rm \,  m^2/K }$ (see, e.g., Hallinan et al. 2017; Mooley et al. 2018). Furthermore, GBT and MeerKAT would potentially be available for the  afterglow observation with $A/T_{\rm sys}\sim 300 {\rm \,  m^2/K }$ (Booth et al. 2009).

For the data analysis   for  SETI, we additionally  have two important parameters. The first one is the coherent  integration time $t_{\rm int}$ of the signal and the second one is the signal bandwidth $B$.   Then, as for an incoming flux density $S$ (usually given in units of Jy$=10^{-23} {\rm erg \,\, s^{-1} cm^{-2} Hz^{-1}}$), the  signal-to-noise ratio (S/N) is  given by
\beq
S/N=\gamma \frac{A\,S\sqrt{B \,t_{\rm int}}}{k_{\rm B}\, T_{\rm sys}} \label{sn}
\eeq
with the Boltzmann's constant $k_{\rm B}$ (Oliver \& Billingham 1971). Here, $\gamma=O(1)$ is a coefficient related to the details of the radio reception system, and we  simply put $\gamma=1$ hereafter.  Conversely, for a given detection threshold $S/N$, the minimum detectable flux density is given by 
\beqa
S&=&4\times 10^{-2} {\rm Jy} \lmk\frac{S/N}{5}  \rmk  \lmk\frac{B}{\rm 0.3 Hz}  \rmk^{-1/2} \lmk\frac{t_{\rm int}}{\rm 10^3 s}  \rmk^{-1/2} \nonumber\\ 
 & & \times  \lmk\frac{A/T_{\rm sys}}{\rm10^4 \,  m^2/K }  \rmk^{-1}. \label{s}
\eeqa
The total energy flux (over the bandwidth $B$) is proportional to $S\cdot B\propto \sqrt{B}$, indicating that the narrower bandwidth is more advantageous for reducing the required power of the ETI (sender), keeping the signal-to-noise ratio $S/N$ at the receivers.  However, the interplanetary scintillation broadens an originally ultra-narrow  line signal, and  the ETI would anticipate this broadening for receivers on  habitable planets.  While the typical broadening at $\nu\sim 1$ GHz is  $\sim 0.01$ Hz on the Earth (Siemion et al. 2015; Harp et al. 2016),  we assume that the ETI  deliver enough energy flux for  $B=0.3$ Hz, including a sufficient margin.  In the next section, we estimate the required transmission power.

As shown in eq. (\ref{s}), we can detect a weaker signal $S$ for a longer integration time $t_{\rm int}$.  However, the required computational resources increase rapidly with $t_{\rm int}$ (Harp et al. 2016). For a SETI data analysis in the SKA era, we set $t_{\rm int}=10^3$ s, presuming  a continuous improvement of available resources.  
Altogether, we would be able to detect the flux density down to $S=0.04$ Jy ($\sim 100$ times smaller than the present value) around $\nu=1$-10 GHz, with the fiducial model parameters given in eq. (\ref{s}). 
 
So far, we have used only the combination $A/T_{\rm sys}$ to characterize a radio telescope. While we do not go into detail, the geometrical configuration of the array could also be important. In  particular, for a host galaxy with a luminous radio source (e.g. AGN), to realize a wide dynamic range, we might need to efficiently control the side lobe pattern of the interferometry.

\section{Transmitter and  Transferable Data Amount}

In this section, we discuss the basic requirements for the transmission facility of the ETI (sender)  and estimate the optimal data amount transferable to a receiver in a  distant Milky Way-like galaxy. To begin  with, we set $d$ as  the spatial dimension of their antenna (or array). Due to the diffraction effects, the beam has the spread angle 
\beq
\alpha=\lmk  \frac{c}{\nu  d}\rmk =6.7\times 10^{-5} {\rm rad} \lmk  \frac{\nu}{3 \rm GHz}\rmk^{-1} \lmk  \frac{d}{1.5 \rm km}\rmk^{-1} .\label{alpha}
\eeq
Below, we use $\nu=3$ GHz and $d=1.5$ km as the fiducial model parameters, both appearing only through the combination $\alpha$.  In reference to \S 2, the angular width $\alpha$ corresponds to the time difference $\sim 2l\alpha/(\pi c)\sim 1$ yr.

To deliver the flux density $S$ to a receiver at a distance $D$  with the bandwidth $B$, the ETI need a total transmission  power
\beqa
P&=& \pi SD^2\alpha^2 B \nonumber      \\ &=& 0.9 {\rm TW} \lmk  \frac{S}{4.0\times 10^{-2}{\rm Jy}}\rmk   \lmk  \frac{D}{\rm 40 Mpc}\rmk^2   \nonumber\\
& & \times \lmk  \frac{\alpha}{6.7\times10^{-5}}\rmk	^2  \lmk  \frac{B}{\rm 0.3 Hz}\rmk.\label{power}
\eeqa
Note that this is not the effective isotropic radiated power but the actual transmission power obtained after multiplying the solid angle $\pi \alpha^2$. For simplicity, we assume that the ETI use the same narrowband transmitter both for the initial alert and the follow-up message, though they might increase the output power for the former (e.g. by decreasing the duty cycle).

To be concrete, let us consider a situation in which the ETI select a target range $D_{\rm max}$. Its likely value  is hard to guess, but here we simply adopt $D_{\rm max}=40$ Mpc, namely the distance between NGC4993 and us.  In addition, they are supposed to take $B=0.3$ Hz (as already mentioned in \S 3), and demand at least $A/T_{\rm sys}=10^4 {\rm m^2/K}$ and $t_{\rm int}=10^3$ s for the technology level of  a receiver (expected in the SKA era). 
Then, from eqs. (\ref{s}) and  (\ref{power}),  the ETI need a transmission power  $P\sim 1$ TW, corresponding to $\sim 10\%$ of the current  energy consumption rate on the Earth (Guillochon \& Loeb  2015). 
In relation to GW170817, with our current technology level,  $A/T_{\rm sys} \sim 300 {\rm \,  m^2/K }$, and $t_{\rm int} \sim 100 $ s, the required power $P$ becomes $\sim$ 100 TW.

Recently, the feasibility of a light-powered propulsion has been actively discussed as a promising technology   for future interplanetary and interstellar missions.\footnote{See, e.g., http://breakthroughinitiatives.org/initiative/3  }
The basic idea is to send a powerful light beam from the Earth to push the sail of a spacecraft.  This can significantly reduce the weight of the spacecraft without onboard fuel after the launch.
Actually, the parameters $d=1.5$ km and $P\sim1$ TW in eqs.(\ref{alpha}) and (\ref{power}) are similar to  the reference values used in  Guillochon \& Loeb (2015) for a potential interplanetary mission (though with $\sim 68$ GHz). Considering the general versatility, we expect that continuous usage of $\sim 1$ TW power would not be unreasonable for an advanced civilization (see also Lingam \& Loeb 2017).

At $D_{\rm max}=40$ Mpc, the emitted  beam width becomes $D_{\rm max}\alpha\sim 2.7$ kpc, somewhat smaller than the angular size of a Milky Way-like galaxy.  Therefore, to cover the whole disk plane ($\sim 10$ kpc radius), they need to divide the corresponding  solid angle into $C\sim 14(D_{\rm max}\alpha/2.7{\rm kpc})^{-2}$ segments.

In the sphere within the distance $D_{\rm max}=40$ Mpc, there would be $N_{\rm T}=4\pi D_{\rm max}^3n_{\rm M}/3\sim 2,700$ target galaxies.\footnote{From the receiver\rq{}s standpoint, the expected available merger rate in the sphere is $N_{\rm T}\eta^{-1}R_{\rm M} \sim 0.1{\rm yr^{-1}}$.}  Given the available sending time $2l/c\sim 7\times 10^4$ years, they can spend $2l/(cN_{\rm T})\sim 20$ yr per galaxy. 
Here, we  assume that the ETI use the square wave and set  the unit duration $t_{\rm int}=10^3$ s for sending 1 bit of information
(almost the same for the frequency-shift keying and polarization-shift keying).  
Taking into account  the segment number $C$, the total data amount transferable to a  receiver in a Milky Way-like disk is roughly estimated to be  
\beq
I\sim \frac{2l}{cN_{\rm T} t_{\rm int} C}\sim 6\times 10^4  {\rm bits}. \label{data}
\eeq
Here, we ignored the overhead for the initial alert signals (mentioned in \S 2).
Considering the segmentations, the data (\ref{data}) might  be received intermittently around  the arrival time of the merger GW signal.

So far, we have fixed the fiducial model parameters $D_{\rm max}$, $\alpha$ and $P$.  Here, still fixing the bandwidth $B$,  we briefly  examine how the data amount changes with these parameters around the fiducial values.  For simplicity, we consider the following scenario; the ETI assume that the system temperature $T_{\rm sys}$ of  a receiver is independent of the frequency (reasonable at least for us in the band  1-10 GHz). From the scaling relations $N_{\rm T}\propto D_{\rm max}^3$, $t_{\rm int}\propto \alpha^{4}P^{-2}D_{\rm max}^{4}$ and $C\propto D_{\rm max}^{-2}\alpha^{-2}$, it follows that 
$I\propto D_{\rm max}^{-5}P^2 \alpha^{-2}$.  Therefore, the transferable data amount (per receiver) decreases rapidly with $D_{\rm max}$, but increases for a smaller beam width $\alpha$.

\section{DISCUSSIONS}

When  searching for an artificial signal from  ETI, a central concern is  how efficiently we can decrease the parameter space under examination.   These circumstances would be inversely understood by the ETI,  and they would carefully arrange the timing and direction of the transmissions. In this paper, we have pointed out that a BNS merger in their galaxy could be an ideal event for the signal synchronization.  This is because the ETI would be able to estimate the location and the epoch of the highly energetic event in advance.  Most optimistically, we might actually find an artificial signal by reanalyzing the electromagnetic data already taken from GW170817. Additionally, the LIGO-Virgo network will start the next observational run in early 2019, and a new BNS merger might be identified. The early and deep radio observation for its host galaxy might also be worth considering from perspective of SETI.\footnote{In the long run, we might empirically identify  a BNS merger only from EM observations.}

As discussed in \S 3 and \S 4, to reach a receiver comparable to the SKA at a distance of $\sim $40 Mpc, the required output power would be $\sim$1 TW.  Such a transmission facility might be realized even on the Earth in 50-100 years. 
Below, we discuss related issues, setting 40 Mpc as the fiducial distance.

So far, we have not mentioned the peculiar velocity of the receiver. For example, relative to the CMB rest frame,  the Milky Way Galaxy is moving at $v\sim 600 {\rm km\, \,  s^{-1}}$, corresponding to the angular deviation $v/c\sim 2\times 10^{-3}$ between the signal transmission and reception. This is much larger than the typical angular size of a target galaxy,  $20 {\rm kpc/40 Mpc}\sim 5\times 10^{-4}$. Therefore, without the correction of the tangential angular velocity, the ETI are likely to miss their shot. If they aim the target galaxy in a scatter-shot fashion, the transferable data amount would become significantly smaller than the optimal value (\ref{data}).  We should stress  that this issue is commonly expected for an extra-Galactic signal transmission, and is not specific to the present synchronization scheme.  It is true that the direct measurement of the tangential  velocity of a distant galaxy would generally be a very difficult task. However, note that the cosmic velocity field has a large-scale coherent component, superimposed by local random velocities (Hoffman et al. 2017).  In this connection, the three-dimensional velocity dispersion in our local group   is $\sim 100 {\rm km\, \,  s^{-1}}$, corresponding to the angular size $\sim 3\times 10^{-4}$ (Courteau \& van den Bergh 1999). If the tangential velocity of a galaxy is estimated with this level, through a successful modeling of its coherent component, the transferable data could be of the same order as  the optimal value (\ref{data}). Note that compared with field galaxies, those in high-density regions (e.g. clusters of galaxies) have  larger random velocities.   In this sense, our Galaxy might be a relatively easy target for extra-Galactic shooters. As a side note, depending on the target distance, other effects (e.g. lensing) might also be relevant.

Eq.(\ref{time}) shows that for $\Delta t_{\rm e}\to 0$, the angle $\theta$ approaches $\pi$.  Correspondingly, viewing from the receiver, the sky position of the ETI moves close to the BNS. Then, due to the confusion, the inherent radio counterpart of the merged BNS  might hamper the identification of the synchronized artificial signal. In  such a case, the ETI would transmit their signal so that it reaches us before the arrival of the merger GW signal (see, e.g., Hallinan et al. 2017; Mooley et al. 2018).  The inspiral wave of GW170817 entered the LIGO band (24 Hz quoted in Abbott et al. 2017a) only $\sim100$ s before the merger.  The Einstein Telescope  is planned to have better sensitivity at the lower-frequency regime down to  $\sim 2$ Hz and could observe the BNS inspiral  $\sim 10^4-10^5$ s before the merger (see eq.(\ref{time})), much earlier than LIGO.
Furthermore, the proposed space detectors BBO and DECIGO will probe the 0.1-10 Hz band (Seto et al. 2001; Cutler \& Holz 2009).
These space detectors will have enough sensitivity to individually resolve the cosmological BNS foreground, for detecting the inflation GW background. Therefore, a BNS at $\sim40$ Mpc would be identified several years before the merger with a firmly localized host galaxy. In other words, BBO/DECIGO will be quite useful for receiving an artificial signal that is delivered designedly before the BNS merger.

The content of the received message would be of great interest.  To clarify the aimed synchronization (not just a  coincidence), the ETI might include the intrinsic information of the BNS. Here, considering the Doppler effect, the mass ratio $q=m_2/m_1$  ($m_2<m_1)$ would be a reasonable choice, rather than the direct mass parameters (e.g. the chirp mass $m_{\rm c}/m_{\rm H}$ normalized by the hydrogen mass $m_{\rm H}$). For GW170817, the LIGO-Virgo network set a very weak constraint $0.7<q<1$ using certain priors for spins (Abbott et al. 2017a). In the future, due to the improved sensitivity and larger dynamic range, the Einstein telescope  and BBO/DECIGO will respectively have $\sim 10^2$ and $\sim 10^4$ better measurement accuracies for the mass ratio $q$. Here, we compared the results in Isoyama et al. (2018) that were given for the reduced mass ratio.  Therefore,   these future projects might also play intriguing roles in decoding a message from ETI.

\acknowledgments
 This work is supported
by JSPS Kakenhi Grants-in-Aid for Scientific Research
(No. 15K65075, 17H06358).

\end{document}